%% file: main.tex
\documentclass[sigconf]{IEEEtran}

\usepackage{booktabs} 

\usepackage{graphicx}
\usepackage{textcomp}
\usepackage{xcolor}
\usepackage{soul}
\usepackage{todonotes}
\usepackage{multirow}
\usepackage{multicol}
\usepackage{float}
\usepackage{cleveref}
\usepackage[nolist]{acronym}
\usepackage{subfigure}
\usepackage[linesnumbered,lined,commentsnumbered,ruled,vlined]{algorithm2e}
\usepackage[]{algpseudocode}
\usepackage{url}
\usepackage[bookmarks=false]{hyperref}

\usepackage[utf8]{inputenc}
\usepackage[english]{babel}
\usepackage{amsmath,amssymb,amsfonts,amsthm}

\usepackage[justification=centering]{caption}

\algtext*{EndWhile}
\algtext*{EndIf}
\algtext*{EndFor}

\DeclareMathOperator{\Ll}{\mathbf{L}}
\DeclareMathOperator{\X}{\mathbf{X}}

\DeclareMathOperator{\Uu}{\mathbf{U}}
\DeclareMathOperator{\UT}{\mathbf{U^{\intercal}}}
\DeclareMathOperator{\g}{\mathbf{g}}
\DeclareMathOperator{\LBD}{\mathbf{\Lambda}}

\begin{document}
\title{Estimating the Circuit De-obfuscation Runtime based on Graph Deep Learning}

\author{
    \IEEEauthorblockN{Zhiqian Chen\IEEEauthorrefmark{1}, Gaurav Kolhe\IEEEauthorrefmark{2}, Setareh Rafatirad\IEEEauthorrefmark{4}, Chang-Tien Lu\IEEEauthorrefmark{1}, Sai Manoj P D\IEEEauthorrefmark{3}, \\Houman Homayoun\IEEEauthorrefmark{2}, Liang Zhao\IEEEauthorrefmark{4}}
    
    \IEEEauthorblockA{\IEEEauthorrefmark{1} Department of Computer Science, Virginia Tech, Blacksburg, VA, USA
    } \\
    \IEEEauthorblockA{\IEEEauthorrefmark{2}
    Department of Electrical and Computer Engineering, University of California, Davis, CA, USA
    } \\
    \IEEEauthorblockA{\IEEEauthorrefmark{3}
    Department of Electrical and Computer Engineering, George Mason University, VA, USA
    }\\
    \IEEEauthorblockA{\IEEEauthorrefmark{4}Department of Information Science and Technology, George Mason University, VA, USA
    \\
\IEEEauthorrefmark{1}\{czq, ctlu\}@vt.edu
\IEEEauthorrefmark{2}\{gskolhe, hhomayoun\}@ucdavis.edu
\{\IEEEauthorrefmark{3}spudukot,
\IEEEauthorrefmark{4}srafatir, \IEEEauthorrefmark{4}lzhao9\}@gmu.edu
}
}

\maketitle
\input{main_body}

\Urlmuskip=0mu plus 1mu\relax
\bibliographystyle{abbrv}
{\scriptsize
\bibliography{main,ref_sai}}

\end{document}

%% file: main_body.tex
\begin{abstract}
Circuit obfuscation has been proposed to protect digital integrated circuits (ICs) from different security threats such as reverse engineering by introducing ambiguity in the circuit, i.e., the addition of the logic gates whose functionality cannot be
determined easily by the attacker. In order to conquer such defenses, techniques such as Boolean satisfiability-checking (SAT)-based attacks were introduced. SAT-attack can potentially decrypt the obfuscated circuits. However, the deobfuscation runtime could have a large span ranging from few milliseconds to a few years or more, depending on the number and location of obfuscated gates, the topology of the obfuscated circuit and obfuscation technique used. To ensure the security of the deployed obfuscation mechanism, it is essential to accurately pre-estimate the deobfuscation time. Thereby one can optimize the deployed defense in order to maximize the deobfuscation runtime. 

However, estimating the deobfuscation runtime is a challenging task due to 1) the complexity and heterogeneity of the graph-structured circuit, 2) the unknown and sophisticated mechanisms of the attackers for deobfuscation, 3) efficiency and scalability requirement in practice. To address the challenges mentioned above, this work proposes the first machine-learning framework that predicts the deobfuscation runtime based on graph deep learning. Specifically, we design a new model, ICNet with new input and convolution layers to characterize the circuit's topology, which is then integrated by composite deep fully-connected layers to obtain the deobfuscation runtime. The proposed ICNet is an end-to-end framework that can automatically extract the determinant features required for deobfuscation runtime prediction. Extensive experiments on standard benchmarks demonstrate its effectiveness and efficiency beyond many competitive baselines.
\end{abstract}

\section{Introduction}

The considerable high capital and operational costs on semiconductor fabrication have motivated most semiconductor companies to outsource it is a fabrication to off-shore foundries. 
Despite the reduced cost and other benefits, 
this trend has led to ever-increasing security risks such as 
IC counterfeiting, piracy and unauthorized overproduction by the contract foundries \cite{subramanyan2015evaluating}. The overall financial risk caused by such counterfeit and unauthorized ICs was estimated to be over \$169 billion per year 
\cite{Doe:2009:Misc}. The major threats from the attackers arise from reverse engineering (RE) an IC and fully identifying its functionality through brute force approaches by applying test inputs and obtaining outputs. 
To prevent such reverse engineering,  \emph{Hardware obfuscation} techniques have been extensively researched in recent years \cite{yasin2017provably}. The general idea is to introduce the ambiguity in the functionality of the IC through obfuscation so that the 
while preserving the original functionality. 
Such techniques were highly effective until the advent of advanced attacking techniques. 
This is based on the fact that there are limited types of gates (e.g., AND, OR, XOR) in IC, so the attackers can just brute force all the possible combinations of types for all obfuscated gates to find out the one that functions identically to the targeted IC to be deobfuscated. As brute force is usually prohibitively expensive,  more recently, efficient methods such as Boolean satisfiability problem (SAT)-based attacks have been proposed,  which have attracted enormous attention \cite{liu2016oracle}. 

The runtime of the SAT attack to reverse engineer 
the IC 
highly depends on the complexity of the obfuscated IC, which can vary from milliseconds to years or more depending on the number and location of obfuscated gates. 
Therefore, a successful obfuscation defence is to increase the amount of time (i.e., many years) required to reverse engineer the design. However, obfuscation comes at a substantial cost in finance, power, and space, and such trade-off requires us to search for optimal positions instead of purely increasing their quantity.
The obfuscation policy tries to select a set of gates such that maximum obfuscation can be achieved while incurring minimal overheads.
Although such selection can significantly influence the deobfuscation runtime, however, until now it is still generally based on human heuristics or experience, which is seriously arbitrary and sub-optimal \cite{khaleghi2018hardware}. This is major because it is unable to ``try and error'' all the different ways of obfuscation, as there are millions of combinations to try and the runtime for each execution (i.e., to run the attacker) can be days, weeks, or years.


To address this issue, this paper focuses on efficient and scalable ways to estimate the runtime of an attacker to reverse engineer an obfuscated IC. This research topic is highly under-explored because of its significant challenges: \textbf{1) Difficulty in characterizing the hidden and sophisticated algorithmic mechanism of attackers.} Over the recent years, a large number of deobfuscation methods have been proposed with various techniques \cite{khaleghi2018hardware}. In order to practically defeat the obfuscation schemes, methods with more and more sophisticated theories, rules, and heuristics have been proposed and adopted. The behaviour of such highly nonlinear and strongly-coupling systems is prohibitive for conventional simple models (e.g., linear regression and support vector machine \cite{bishop2014pattern}) to characterize. \textbf{2) Challenge in extracting determinant features from discrete and graph-structured IC.} The inputs of the runtime estimation problem is the IC and the selected gates for obfuscation, where the first input is a heterogeneous graph while the second is a vector with discrete values. Conventional feature extraction methods are not intuitive to be applied to such type of data without significant information loss. Hence, it is highly challenging to instantly formulate and seamlessly integrate them as mathematical forms that can be input to conventional computational and machine learning models. \textbf{3) Requirement on high efficiency and scalability for deobfuscation runtime estimation.} The key to the defence against deobfuscation is the speed. The faster the defender can estimate the deobfuscation runtime for each candidate set of obfuscated gates, the more candidate sets the defender can evaluate, and hence the better the obfuscation effect will be. Moreover, the estimation speed of deobfuscation runtime must not be sensitive to different obfuscation strategies in order to make the defender strategy controllable.

This work addresses all the above challenges and proposes the first generic framework for deobfuscation runtime prediction, based on graph deep learning techniques. In the recent years, deep learning methods in complex cognitive tasks such as object recognition and machine translation have achieved immense success \cite{Wang'19,Lechner_IGSC'19}, which motivates the generalization of it into graph-structured data \cite{kipf2017semi}. By concretely formulating ICs and the obfuscated gates as multi-attributed graphs, this work innovatively leverages and extends the state-of-the-art graph deep learning methods such as Graph Convolutional Neural Networks (GCN) \cite{kipf2017semi} to instantiate a graph regressor. Such end-to-end deep graph regressor can characterize the underlying and sophisticated cognitive process of the attacker for deobfuscating the ICs. To adopt the powerfulness of GCN and handle the aforementioned issues, we extend it by adjusting the connectivity representation inspired by domain facts. Our enhanced GCN can automatically extract the discriminative features that are determinants to the estimation of the deobfuscation runtime to achieve accurate runtime prediction. After being trained, the prediction based on this deobfuscation runtime estimator just runs instantly fast by simply performing a feed-forward propagation process. 
The major contributions of this paper are:
\begin{itemize}
    \item Proposing a new framework, ICNet, for deobfuscation runtime estimation based on graph deep learning.
    \item Developing a new multi-attributed graph convolutional neural network for graph regression.
    \item Conducting systematical experimental evaluations and analyses on real-world datasets (ISCAS-85 benchmark).
\end{itemize}
The rest of the paper is organized as follows. Section \ref{sec:related_work} reviews the existing work. Section \ref{sec:method} elaborates proposed graph learning model for SAT runtime prediction. In Section \ref{sec:evaluation}, experiments on real-world data are presented. This paper concludes by summarizing the study's important findings in Section \ref{sec:conclusion}.

\vspace{-5pt}
\section{Background and Related Work}\label{sec:related_work}
We discuss the logic obfuscation and SAT attacks followed by graph convolutional networks and the relevant works. 

\vspace{-5pt}
\subsection{Logic Obfuscation and SAT Attacks}
Logic obfuscation often referred to as logic locking \cite{logiclock} is a hardware security solution that facilitates to hide the IP 
using key-programmable logic gates 
The activation of the obfuscated IP 
is accomplished in a trusted regime before releasing the product into the market, thereby reducing the probability to obtain the secret configuration keys by the attacker. During the activation phase, the correct key is applied to these key-programmable gates to recover the correct functionality of the IC/IP. Besides, the correct key will be stored in the IC in a tamper-proof memory. 
Although obfuscation schemes try to minimize the probability of determining the correct key by an attacker, thereby curbing the ongoing piracy of the legitimate IPs. However, SAT attack shows that the contemporary obfuscation schemes can be broken \cite{Subramanyan} to retrieve the correct key. In order to perform 
SAT attack, the attacker is required to have 
The 
SAT attack first tries to find the Distinguishing Input Patterns (DIP) X${_i}$, 
which when applied as the input can produce different outputs ($Y_i$) 
such that ($Y_1 \neq Y_2$)
when different key values are applied (K${_1}$, K${_2}$). This DIP can then be used to distinguish the correct and incorrect keys. The number of DIPs discovered during the SAT-based attack is the same as the number of iterations needed to unlock the obfuscated design. 
In each iteration, a constraint is added to SAT solver, until SAT solver cannot find a satisfying assignment.
This results in finding the correct key. 


    
Different SAT-hard schemes such as \cite{Kolhe:2019:CLO:3299874.3319496,iccad:2019:CLO:3299874.3319496} are proposed 
Furthermore, new obfuscation schemes that focus on non-Boolean 
Behaviour of circuits \cite{delay_lock}, that are not convertible to an SAT 
circuit is proposed for SAT resilience. Some of such defences include adding cycles into the design \cite{SRCLock}. By adding cycles into the design may cause that the SAT attack gets stuck in the infinite loop, however, advanced SAT-based attacks such as cycSAT \cite{cycSAT} can extract the correct key despite employing such defences. 
To ensure that the proposed defence ensures robustness against SAT attacks, the defenders need to run the rigorous simulations which could range from few minutes up to a few days. Furthermore, this can be exacerbated when the defender verifies for large and real-world circuits. This work proposes the use of graph convolutional networks (GCNs) 
to alleviate the need to run the attack to verify whether the defence is strong enough or not. 
The work in \cite{Neurosat} utilizes neural network with single-bit supervision to predict whether a given circuit in Conjunctive Normal Form (CNF) can be decrypted or not. However, this is limited to determining for few kinds of SAT-solvers, but cannot be applied to SAT-hard solutions such as SMT-SAT \cite{kimia'19}, a superset of SAT attacks.
However, with the proposed graph convolutional network (GCN) based predictor, the defender can determine the deobfuscation time in a single run of GCN, which consumes a few seconds. 
We introduce the GCN below. 

\vspace{-12pt}
\subsection{Graph Convolutional Networks}
Spectral graph theory is the study of the properties of a graph in relationship to the characteristic polynomial, eigenvalues, and eigenvectors of matrices associated with the graph. Many graphs and geometric convolution methods have been proposed recently. The spectral convolution methods \cite{defferrard2016convolutional,kipf2017semi} are the mainstream algorithms developed as the graph convolution methods. 
Their theory is based on the graph Fourier analysis \cite{shuman2013emerging}. The polynomial approximation is firstly proposed by \cite{hammond2011wavelets}. Inspired by this, graph convolutional neural networks (GCNs) (\cite{defferrard2016convolutional}) is a successful attempt at generalizing the powerful convolutional neural networks (CNNs) in dealing with Euclidean data to modelling graph-structured data. Kipf and Welling proposed a simplified type of GCNs \cite{kipf2017semi}, called graph convolutional networks (GCNs). The GCN model naturally integrates the connectivity patterns and feature attributes of graph-structured data and outperforms many state-of-the-art methods significantly. 


Therefore, it is promising to apply GNNs for the circuit problem, since ICs can be naturally represented using a graph with connectivity among gates.


\section{Proposed Model for Runtime Prediction}\label{sec:method}
This section introduces the problem setting, and we present the deobfuscation time prediction through the proposed ICNet.


\vspace{-10pt}
\subsection{Problem Setting}
First, a circuit is modeled as a graph network: $\mathcal{G} = (\mathcal{V}, \mathcal{E}, \mathcal{W})$, where $\mathcal{V}$ is a set of $n$ vertexes (gates), $\mathcal{E}$ represents links among gates and
$\mathcal{W} = [w_{ij}] \in \{0,1\}^{n\times n}$ is an unweighted adjacency matrix. 
A signal $\X$ defined on the nodes is regarded as a vector $\X \in \mathbb{R}^{n\times F}$. A graph structure representation, i.e., combinatorial graph Laplacian, is defined as $\Ll= D-\mathcal{W} \in \mathbb{R}^{n\times n}$ where $D$ is degree matrix.  
Accordingly, we formulate the estimation of running time on IC as a regression task. Specifically, the model accepts graph structure along with gate features as input, and predict the running time:
\begin{equation}
\small
Y=f(\mathcal{G}, \X)\Theta,
\label{eq:problem}
\end{equation}where $f$ is a function integrating graph structure $\mathcal{G}$ and gate feature $\X$. $\mathcal{G}$ is often represented by graph Laplacian $\Ll$ in graph theory \cite{chung1997spectral}
$\Theta$ indicates the parameters of fully neural network layers connecting the actual runtime $Y$ and $f$. The purpose of $\Theta$ is (1) fitting dimension with $Y$ and (2) generalizing the logic pattern between $Y$ and $f$.
The goal of \ref{eq:problem} is to learn $f$ and $\Theta$ so that the difference between $Y$ and $f(\mathcal{G}, \X)\Theta$ is minimized.
However, there exists no straightforward relationship among the number of obfuscated gates, type of obfuscated gates and other factors to determine the deobfuscation time. This makes it harder to efficiently estimate the SAT-attack runtime with traditional machine learning models. 
Hence, we survey a thread of works called graph neural networks or geometric deep learning to address the problem of deobfuscation estimation, as the netlist can be perceived as the graph representation of various logical elements. 



\begin{figure}[!htb]
\small
\centering
	\includegraphics[width=0.32\textwidth]{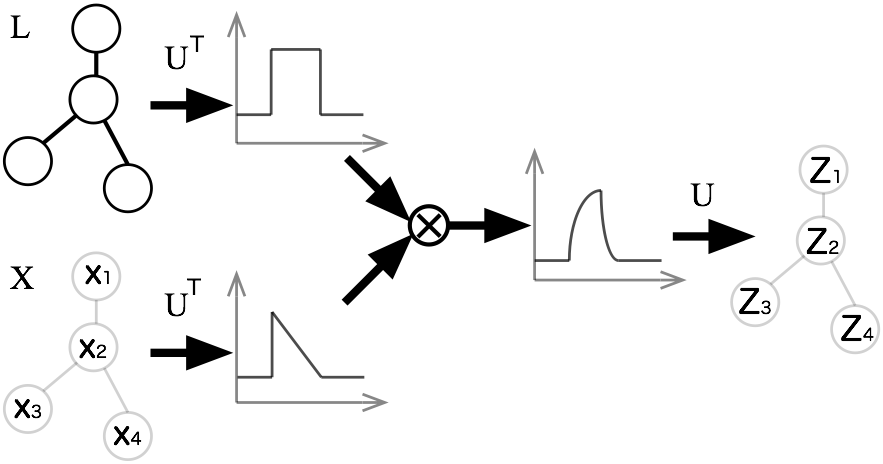}
	\caption{GCN workflow.}
	\label{fig:gcn_example1}
\end{figure}
\vspace{-10pt}

Graph convolutional network (GCN) is a recently emerging technique that integrate graph structure and node attributes, and its general process is performed as follows: 
it determines a vertex complete set of orthonormal Eigen vectors (frequency components) and their associated ordered real non-negative eigenvalues identified as the weights of these frequencies components. Specifically, the Laplacian is first diagonalized by the Fourier basis $\UT$: $\Ll = \Uu \Lambda \UT$ where $\Lambda$ is  the diagonal matrix whose diagonal elements are the corresponding eigenvalues, i.e., ${\displaystyle \Lambda _{ii}=\lambda _{i}}$. The graph Fourier transform of a signal $\X\in \mathbb{R}^{n\times F}$ is defined as $\hat{\X}=\UT \X \in \mathbb{R}^{n}$ and its inverse as $\X=\Uu \hat{\X}$ \cite{shuman2013emerging,shuman2016vertex}. To enable the formulation of fundamental operations such as filtering in the vertex domain, the convolution operator on graph is defined in the Fourier domain such that $f_{1}*f_{2}=\Uu \left[\left(\UT f_{1} \right) \otimes \left(\UT f_{2}\right)\right]$, where $\otimes$ is the element-wise product, and $f_{1}/f_{2}$ are two signals defined on vertex domain. The intuitive workflow of GCN is shown in Figure \ref{fig:gcn_example1}. It follows that a vertex
signal $f_{2}=\X$ (gate features) is filtered by spectral signal $\hat{f_{1}}=\UT f_{1}=\g$ (graph structure) as:
\begin{equation*}
\scriptsize
\g * \X = \Uu \left[\g(\LBD)\odot \left(\UT f_{2}\right)\right] = \Uu \g(\LBD) \UT \X.
\end{equation*}
\vspace{-10pt}

\vspace{-10pt}
\subsection{Proposed Model: ICNet}
Our proposed method, namely ICNet, is a neural network that is based on graph convolution operator. As shown in Figure \ref{fig:icnet}, ICNet encodes the obfuscated circuit into two components:
\begin{figure*}[!hbtp]
\centering
  \includegraphics[width=6.6in]{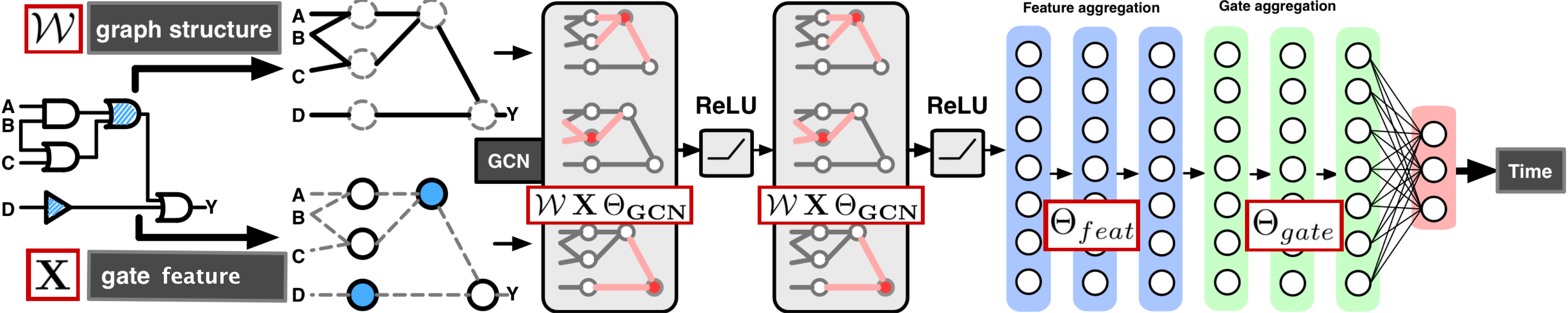}
  \caption{Illustration of ICNet structure: Two graph convolutions ($\mathcal{W}\X\Theta_{\mathbf{GCN}}$) followed by ReLU activation, and attention layers for features ($\Theta_{feat}$) and gate ($\Theta_{gate}$) respectively.}
  \label{fig:icnet}
 \vspace{-20pt}
\end{figure*}
\begin{itemize}
  \item \textbf{Graph Structure $\mathcal{G}$}: Complete set of local connections are often used to represent the graph structure \cite{chung1997spectral}. Typically, a graph Laplacian is employed, in this work since it contains gate-wise connections.
  \item \textbf{gate features $\X$}: Gate-level information is encoded as numerical vector as input feature. Such information could include gate type, whether it is obfuscated and so on.
\end{itemize}
By applying the GCN, we can easily build a model to learn the relationship between the circuit and deobfuscation time automatically. However, GCN suffers from several issues: \textbf{(1)} the original graph convolutional operator is not suitable for the circuit since the graph Laplacian will make the graph convolutional operator behaviour as label propagation, i.e., the attributes of each gate are similar to its neighbours. This is called the smoothness assumption \cite{li2018deeper}, and it does not fit the fact
that gate type or encryption location of each gate does not determine its neighbours' related attributes in theory.
This issue is due to that graph Laplacian matrix is used during graph convolution operation, which counts each node as -$N_{i}$ ($i$ is the index of the row in graph Laplacian), and counts the weighted sum of its neighbours as $N_{i}$. Consequently, they are cancelled out when gate representation are aggregated using sum, and the model can hardly learn the relationship between their sum (residues) and actual runtime. 
\textbf{(2)} The default setting of GCN aggregate gates and their features using the mean function, which is not supported by any domain knowledge, and not likely to cover the actual pattern on features or gates.
To solve these issues, our model employs several policies to enhance the traditional GCN for circuit learning. 

\begin{itemize}
  \item \textbf{Graph Representation $\mathcal{G}=A$}: Our model uses adjacency matrix $A$ instead of graph Laplacian. This representation can avoid intrinsic smoothness assumption which is not compatible with ICs.
  \item \textbf{Feature Aggregation($\Theta_{feat}$)}: The mean function is a typical methods for aggregating node feature. However, the mean function does not consider the quantity of sum. A more flexible way is to learn feature aggregation by a neural network automatically.
  \item \textbf{Gate Aggregation($\Theta_{gate}$)}: similarly, mean function can also be used to aggregate gate representation. Due to the complicated real-world aggregation, another neural network is designed to learn the gate aggregation function for more flexibility.
\end{itemize}

Our model is based on GCN 
which simplify the layer parameters of graph convolutional operator and applies an approximate technique to boost the efficiency. GCNs, as a state-of-the-art deep learning method for the graph, focus on processing graph signals defined on undirected graphs.
According to the analysis above, graph Laplacian is replaced with adjacency matrix. To fit whole-graph level regression task, the proposed method designs two aggregation neural networks. Formally, it is denoted as:
\begin{equation}
\begin{aligned}
Y=&f(\mathcal{G}, \X) \Theta &&{\scriptstyle{\text{(GNN definition Eq. \ref{eq:problem})}}}\\
=& \underbrace{\mathbf{GCN}(\mathcal{W}, \X)}_\text{f=GCN}\Theta &&{\scriptstyle{\text{(apply GCN with adjacency matrix)}}}  \\
=& \mathbf{GCN}(\mathcal{W}, \X)\Theta_{feat}\Theta_{gate}&&{\scriptstyle{(\Theta\rightarrow\{\Theta_{feat}, \Theta_{gate}\})}}  \\
= & \underbrace{\sigma(\mathcal{W}\X\Theta_{GCN}}_\text{GCN})\Theta_{feat}\Theta_{gate},
&&{\scriptstyle{\text{(rewrite GCN in matrix form)}}} 
\end{aligned}    
\end{equation}
where activation $\sigma$ is implemented by ReLU function. The running time tends to grow at an exponential rate as the number of encrypted gates increase. Therefore, the model is modified as:
\begin{align}
Y = \exp({\mathcal{W}\X\Theta_{\mathbf{GCN}}\Theta_{feat}\Theta_{gate}})
\label{eq:icnet}
\end{align}

As illustrated in Fig. \ref{fig:icnet}, the proposed ICNet conducts two graph convolutional operations (GCN) to fuse the information from graph structure and gate features. Then two sets of neural networks are performed for the feature and gate aggregation. To further increase the model's interpretability, we replace fully connected layers $\Theta_{feat}$ and $\Theta_{gate}$ as follows:
Generally, \textit{sum} or \textit{mean} function is a typical method for aggregating node attribute into lower dimensional vector. This  treats voting from each gate equally, which is not fit in theory. For example, the encrypted gate should be weighed higher, since it impose more difficulty on obfuscation task; gate types also have a significant impact on runtime \cite{subramanyan2015evaluating}. Therefore, a more flexible way is to build  a neural network to automatically learn attribute aggregation. To fit the whole-graph level regression task, the proposed method designs two aggregation neural components based on soft attention mechanism \cite{velivckovic2018graph} for feature-level and gate-level. Formally, the feature based attention is calculated as:
{
\begin{equation}
\begin{aligned}
a_{i}=\frac{exp(e_{i})}{\sum_{i} exp(e_{i})}, e_{i}=\sum_{i} \theta_{i}\mathbf{F}_{i},
\label{attention}
\end{aligned}    
\end{equation}}where $\mathbf{F}_{i}$ represents $i$th feature after $\mathbf{GCN}$, $\theta_{i}$ is the weight parameter for $\mathbf{F}_{i}$, $a_{i}$ is the corresponding attention and thereby the output of this layer is $\sum_{i}a_{i}\mathbf{F}_{i}$. This attention shows which feature contributes more to the obfuscation time. Similarly, gate-wise attention is utilized for gate-level aggregation by setting $\mathbf{F}_{i}$ to $i$th gate in \eqref{attention}. 
\vspace{-10pt}
\begin{algorithm}[!h]
    \small
    \caption{ICNet}
    \label{algo:fgan}
    \SetAlgoLined
    \KwIn{An integrated circuit graph $\mathcal{G}=\{\mathcal{V}, \mathcal{E}\}$, gate features set: $x_{j}(i)$, $i$ $\in$ {$1, 2, ..., |\mathcal{V}|$} for each encryption instance $D_{j}$, the real runtime $Y_{j}$ for instance $D_{j}$}
    \KwOut{A neural network function with parameters $\Theta_{\mathbf{GCN}}$, $\Theta_{feat}$ and $\Theta_{gate}$}
    // Data preparing \\
    Calculate $\mathcal{W}$ which is the adjacency matrix of $\mathcal{G}$ \\
    Split encryption instances $D$ into training set $D_{train}$ and testing set $D_{test}$ \\
    Split both $D_{train}$ and testing set $D_{test}$ into batch set $d_{train}$ and testing set $d_{test}$ \\
    // Update ICNet\\
    $\theta=\{\Theta_{\mathbf{GCN}}, \Theta_{feat},\Theta_{gate}\}$\\
    Initialize $\theta$ with Gaussian or uniform distribution. \\
    \Repeat{$\delta$ convergence}{
        Randomly select one $d_{train}={x_{d1}, x_{d2}, ...}$ \\
        Calculate predicted runtime $\hat{Y}$   \Comment{Eq. \ref{eq:icnet} and \ref{attention} }\\
        Calculate residues $\delta= Y-\hat{Y} $ \\
        Compute derivatives to update parameters: $ \theta \leftarrow \theta + \beta\nabla_{\theta}\delta$, where $\beta$ is learning rate
    }

\end{algorithm}
\vspace{-6pt}

\vspace{-6pt}
\subsection{Algorithm description}
The Algorithm \ref{algo:fgan} first prepare graph adjacency as circuit connection representation (line 2). To fit the machine learning schema, the whole dataset is split into training and testing dataset. Each dataset is then split into small batch size to improve learning efficiency (line 3-4). ICNet training is an iterative process which updates the model until the residues are small enough or converged (line 6-13). First, the model parameters are initialized by Gaussian or uniform distribution. In each iteration, a batch of the training set is selected randomly. By equation \ref{eq:icnet}, the model computes the predicted runtime (line 10) and then calculates the residues between real runtime and prediction (line 11). Following normal deep learning schema, the model update parameters by the derivatives regarding the parameters themselves with learning rate (line 12).

\vspace{-10pt}
\section{Evaluation}\label{sec:evaluation}
This section elaborates evaluation of the proposed method ICNet with competitive baselines including:
Graph deep learning methods: GCN \cite{kipf2017semi}, ChebNet  \cite{defferrard2016convolutional}.
The input of these models above is exactly same as our model. We also compare against several state-of-the-art regression models\footnote{https://scikit-learn.org/stable/modules/linear\_model.html}:
Linear Regression (LR), LASSO \cite{tibshirani1996regression}, Epsilon-Support Vector Regression(SVR) (Two kernels were applied:  polynomial (P) and RBF (R)), \cite{smola2004tutorial}, Ridge Regression (RR) \cite{ng2004feature}, Elastic Net (EN) \cite{zou2005regularization},  Orthogonal Matching Pursuit (OMP) \cite{mallat1993matching}, SGD Regression, Least Angle Regression (LARS) \cite{efron2004least}, Theil-Sen Estimators (Theil) \cite{dang2008theil}. These regression models does not model graph using Laplacian or adjacency matrix, since they can only accept feature vector. Therefore, the input are encoded as mean or sum on concatenation of Laplacian or adjacency matrix and gate features. 

\vspace{-10pt}\subsection{Data processing}
The datasets are obtained by running SAT algorithm \cite{Subramanyan,subramanyan2015evaluating} on real-world ISCAS-85 benchmark: First, we take a circuit and select a random gate and replace it with LUT of fixed size (LUT size 4 in current work). To deobfuscate, we implement SAT attack algorithm \cite{Subramanyan,subramanyan2015evaluating} with the obfuscated circuit netlist as input. We monitor the time that SAT-attack takes to decode the key, which is the deobfuscation time. The proposed model is evaluated on two datasets: 
\textbf{Dataset 1}: the total number of the encryption location ranges from 1 to 350, this is for testing if the model is sensitive to the number of encrypted quantity of gates. \textbf{Dataset 2}: the total number of the encryption location ranges from 1 to 3, this is for testing if the model can handle very small value.

The circuit in the experiments is the same, and the total gate number of the circuit is 1529. For graph deep learning methods, the graph is represented using Laplacian matrix or adjacency matrix, while for general regression baselines, the graph Laplacian or adjacency matrix is summed or averaged across gates. 
Though the evaluations showed here are mere proof-of-concept of how powerful the proposed GCN based deobfuscation runtime prediction is, it can be applied to an SAT-hardening solution utilizing any replacement policy, LUT size and other SAT parameters, by retraining GCN.
\vspace{-8pt}
\subsection{Experiment configuration}
The features of gate used in experiments include: \textbf{gate mask}: if the gate is encrypted, the value is set to 1, otherwise 0. and \textbf{gate type}: the gate type include  \{AND, NOR, NOT, NAND, OR, XOR\}, they are encoded using one-hot coding, such as [1,0,0,0,0,0] for AND and [0,1,0,0,0,0] for NOR gate.

For graph deep learning model (ChebNet and ICNet), the graph structure is represented using graph Laplacian matrix or adjacency matrix. These model employ ADAM \cite{le2011optimization} optimizer and will stop learning when the learning loss is converged. The implementation of our model will be available online. All the baselines and the proposed model are tested on two different feature set, since gate type is useful or not is unknown.: \textbf{Location}: Only the gate mask is included. \textbf{All features}: Besides gate mask, gate type is also included.

For node aggregation, we apply $sum$, and $mean$ since they are popular. Deep learning model can have another node aggregation method, i.e., learning by a neural network automatically. Therefore, in the results, ChebNet-NN and ICNet-NN denote the automatic version. It is expected that a deep neural network can learn an optimal aggregation which is not worse than our assumption, i.e., sum or mean.

\vspace{-5pt}
\begin{table}[hbt]
    \centering
    \caption{Regression Performance (MSE) on Dataset 1}
    \label{data1}
    \scriptsize
    \begin{tabular}{l|cc|cc}
        \toprule \hline
        & \multicolumn{2}{|c|}{Location} &\multicolumn{2}{c}{All feat}\\
        \hline
        Method & Sum &  Mean & Sum &  Mean  \\ \hline
SVR RBF & 1.6791   & 0.6784 & 1.6675 & 0.6739\\
SVR Poly & 0.1913  & 2.1890  & 0.1696 & 2.2091\\
SGD & 2.1450e+25 & 2.1823 & 1.0430e+26 & 2.2072 \\
LR & 0.2839 & 0.2284 & 0.2449  & 0.2253\\
RR & 0.2309 & 2.1508 & 0.2058  & 2.1738\\
LASSO & 0.9213  & 2.1843 & 1.0127 & 2.2083\\
EN & 0.5763 & 2.1843 & 0.6409 & 2.2083 \\
OMP & 1.8182 & 1.9192 & 1.8651  & 2.0337 \\
LARS & 1.9968 & 2.1277 & 2.0434 & 2.1833\\
Theil & 0.2948   & 0.2238 & 0.2385 & 0.2277\\
\hline
ChebNet &  0.1484   &  8.8370e+33 & 0.1761 & 0.1760 \\
ChebNet-NN & \multicolumn{2}{|c|}{0.17858} & \multicolumn{2}{|c}{3.8549e+27} \\
GCN &  0.3364   &  0.4149 & 0.2496 & 0.3290 \\
GCN-NN & \multicolumn{2}{|c|}{0.1811} & \multicolumn{2}{|c}{0.1606} \\
ICNet & 0.1534   & 0.1256  &  0.2390 & 0.1902 \\
ICNet-NN & \multicolumn{2}{|c|}{\textbf{0.0843}} &  \multicolumn{2}{|c}{\textbf{0.1367}}\\
\hline
\bottomrule
    \end{tabular}
\end{table}
\vspace{-10pt}

\vspace{-10pt}
\subsection{Regression Results}
In the dataset 1 experiment (Table \ref{data1}, all methods achieved acceptable mean square error (MSE) except SGD (sum) which did not learn a reasonable model to predict the runtime, since the value is tremendous (at e+25/+26 scale). Most regression methods are sensitive to the aggregation method. For example, only using location feature, MSE 
of RR is 0.2309 when using sum, but it got 2.1508 when using the mean function. Sensitive models include SVR, LASSO, and EN. The best of the regression baselines is SVR (ploy), which achieved MSE of 0.1913.
On the other hand, ChebNet is slightly better than the best regression model. However, ChebNet is not stable and sensitive to the aggregation method and feature set since it may yield a substantial error. Our proposed ICNet-NN
is stable to the feature and aggregation setting and outperformed all the other methods, i.e., 0.0843 of MSE. Note that ICNet-NN is better than ICNet with sum or mean function, which demonstrates that there exists a better aggregation method, and graph neural network can learn it automatically. ICNet is always better than GCN under any settings, which shows that our improvement based on GCN works on circuit scenario.
\begin{table}[hbt]
    \centering
    \scriptsize
        \caption{Regression Performance (MSE) on Dataset 2}
    \label{data2}
    \begin{tabular}{l|cc|cc}
        \toprule \hline
        & \multicolumn{2}{|c|}{Location} &\multicolumn{2}{c}{All feat}\\
        \hline
        Method & Sum &  Mean & Sum &  Mean  \\ \hline
SVR RBF & 0.0051 & 0.0048 & 0.0050 & 0.0051\\
SVR Poly & 0.0048 & 0.0048 & 0.0048 & 0.0051 \\
SGD & 7.6301e+25 & 0.0045 & 2.0675e+26 & 0.0049 \\
LR & 6.9063e+23 & 4.6521e+20 & 7.2916e+25 & 5.8600e+23\\
RR & 0.0070 & 0.0045 & 0.0065 & 0.0049\\
LASSO & 0.0047 & 0.0045  & 0.0046 & 0.0049\\
EN & 0.0047 & 0.0045 & 0.0046 & 0.0049 \\
OMP & 0.0047 & 0.0045 &  0.0045 & 0.0049 \\
PAR & 0.0054 & 0.1918 & 0.0051 & 0.3143\\
LARS & 0.0047 & 0.0045 & 0.0046 & 0.0049\\
Theil & N/A & N/A & N/A & N/A \\
\hline
ChebNet &  0.0047   &  0.0045 & 4.3570e+28 & 0.0048 \\
ChebNet-NN & \multicolumn{2}{|c|}{\textbf{0.0043}} & \multicolumn{2}{|c}{0.0047} \\
GCN &  0.0061   &  0.0046 & 0.0048 & 0.0050 \\
GCN-NN & \multicolumn{2}{|c|}{0.0050} & \multicolumn{2}{|c}{0.1606} \\
ICNet & 0.0049   & 0.0047  &  \textbf{0.0040} & 0.0043 \\
ICNet-NN & \multicolumn{2}{|c|}{0.0051} &  \multicolumn{2}{|c}{0.0048}\\
\hline
\bottomrule
    \end{tabular}
\end{table}
While in the dataset 2 (Table \ref{data2}), it is more challenging, since all the runtime are small and the model has to be very precise to achieve low MSE. All methods at almost the same level of MSE. Once again, some of the regression models are not stable such as SGD and LR. Graph deep learning method includes ChebNet and ICNet still at the best error level. ChebNet can achieve the best level but sensitive to the settings (i.e., location or all feature), while ICNet is insensitive to this setting. Therefore, ICNet is more stable than GCN and ChebNet, since the difference of MSE between location and all feature is smaller than that of ChebNet or GCN. Under all feature setting, ICNet-NN is still the best method, and it outperformed its mean and sum version. 

The proposed method, ICNet, not only predicted the value very precisely but also with small variance. The runtime of ICNet is 1.1336 seconds on average, ranging from 1 to 2 seconds on dataset 1 and 2. This is because runtime of ICNet only depends on its parameter number. The instance with the largest runtime on dataset 1 and 2 spends 2411.11 seconds by actual solver. Therefore, ICNet can save 99.95\% of solver's time to get accurate runtime.


\begin{figure*}[t!]
\centering
\subfigure[EN]{\includegraphics[width=1.36in]{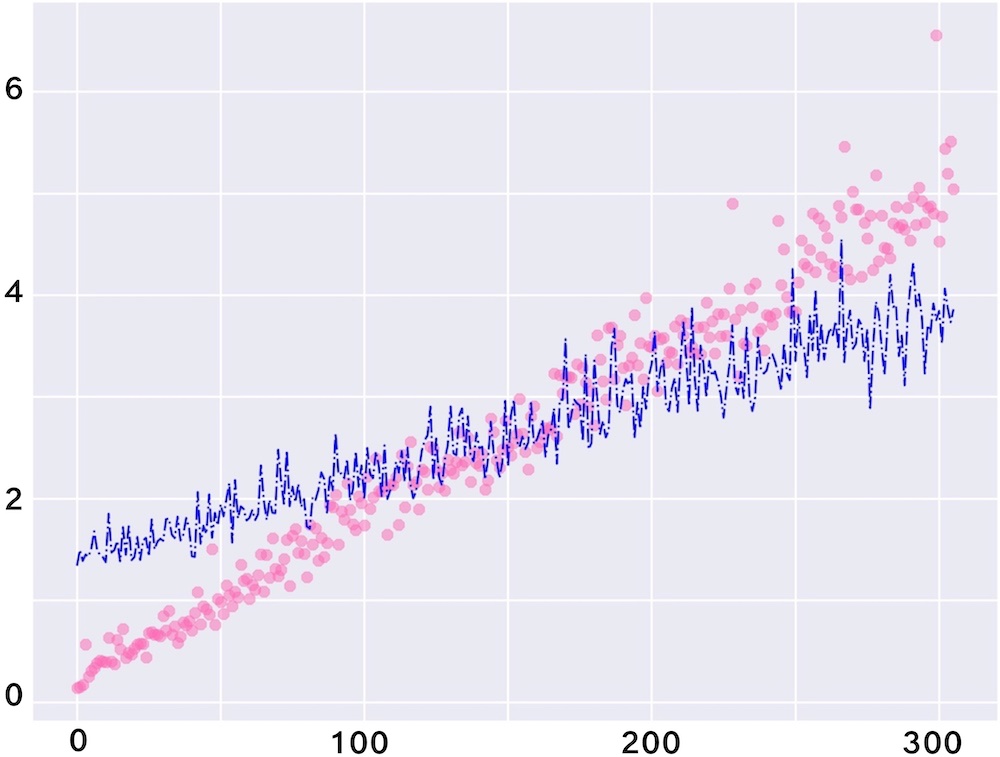}}
\subfigure[LASSO]{\includegraphics[width=1.36in]{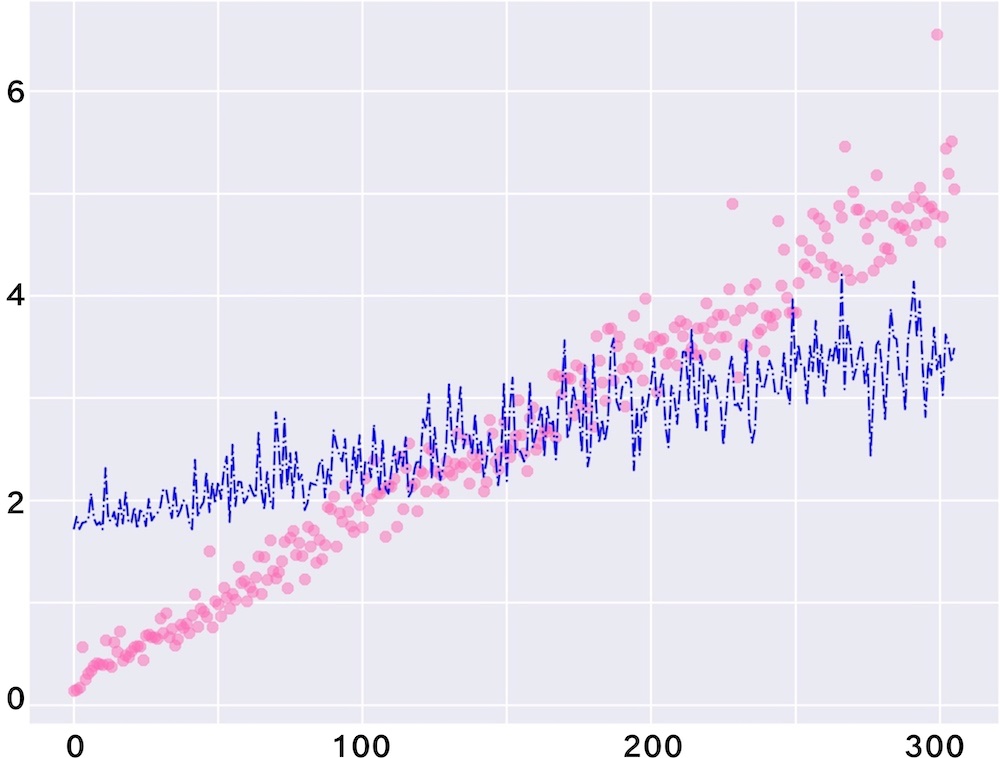}}
\subfigure[Linear]{\includegraphics[width=1.36in]{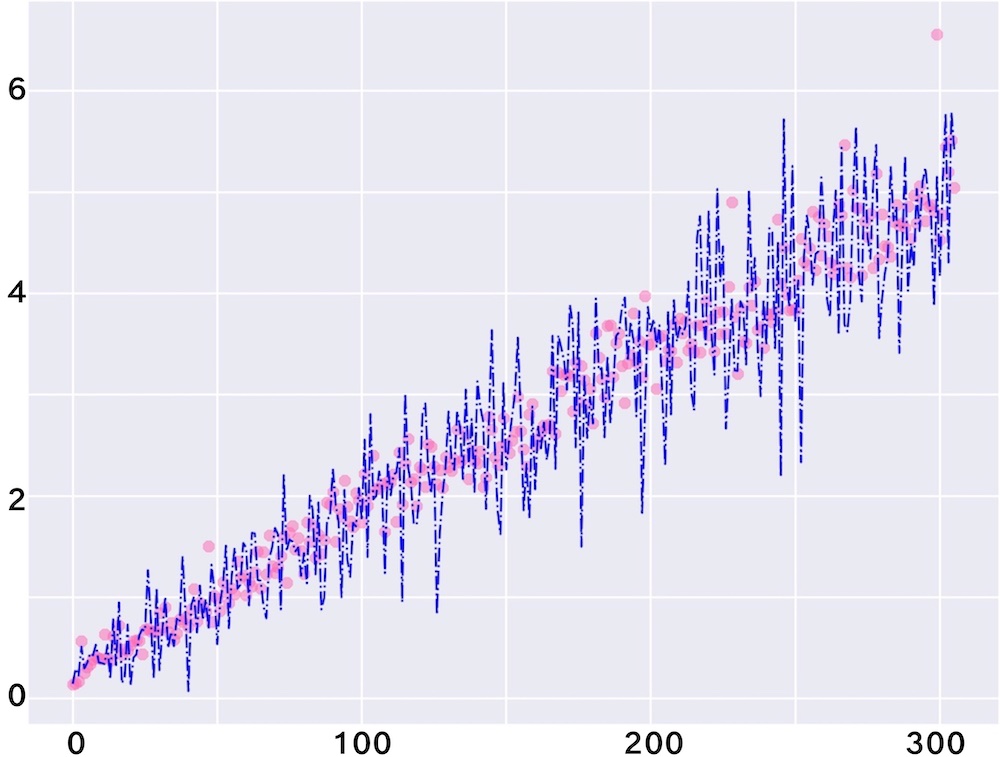}}
\subfigure[OMP]{\includegraphics[width=1.36in]{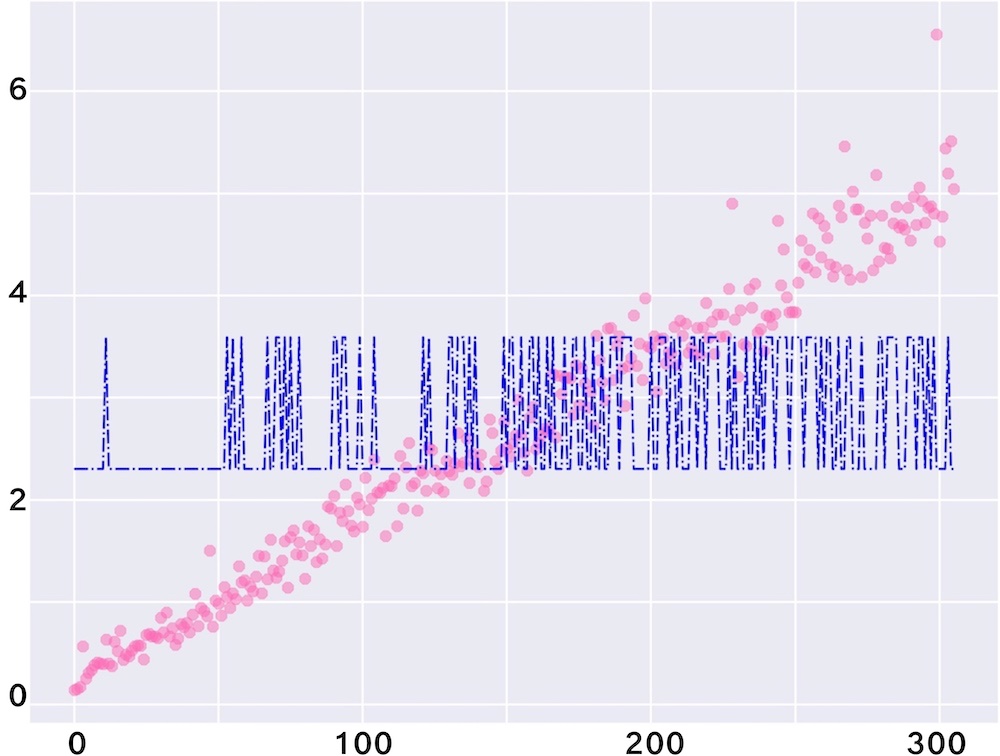}}
\subfigure[RR]{\includegraphics[width=1.36in]{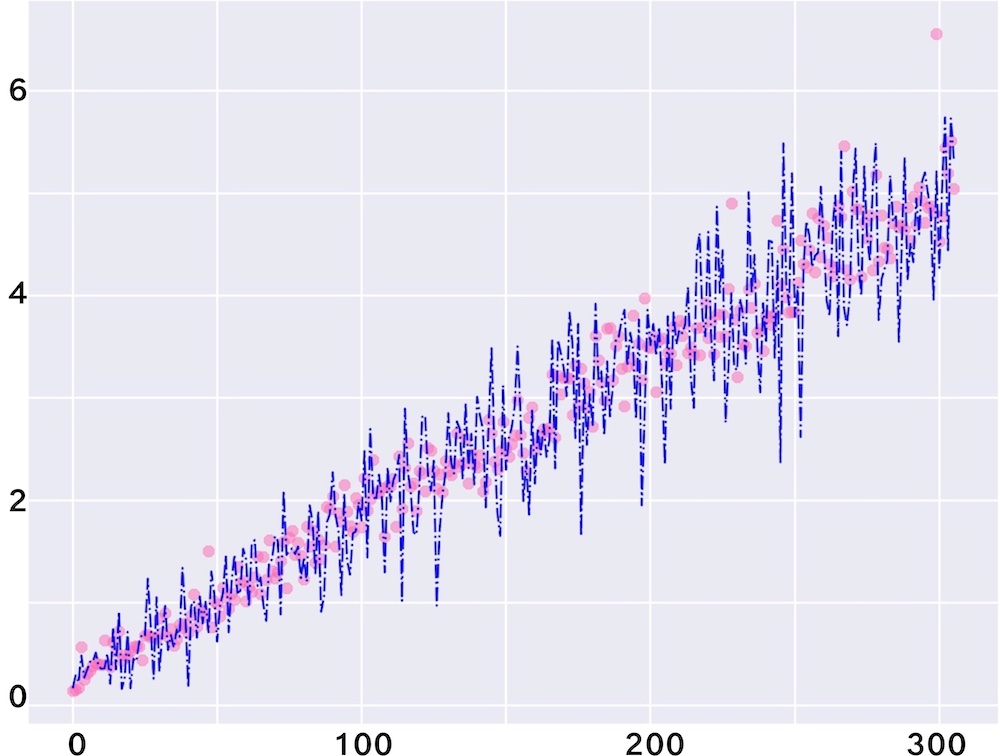}}

\subfigure[SGD]{\includegraphics[width=1.36in]{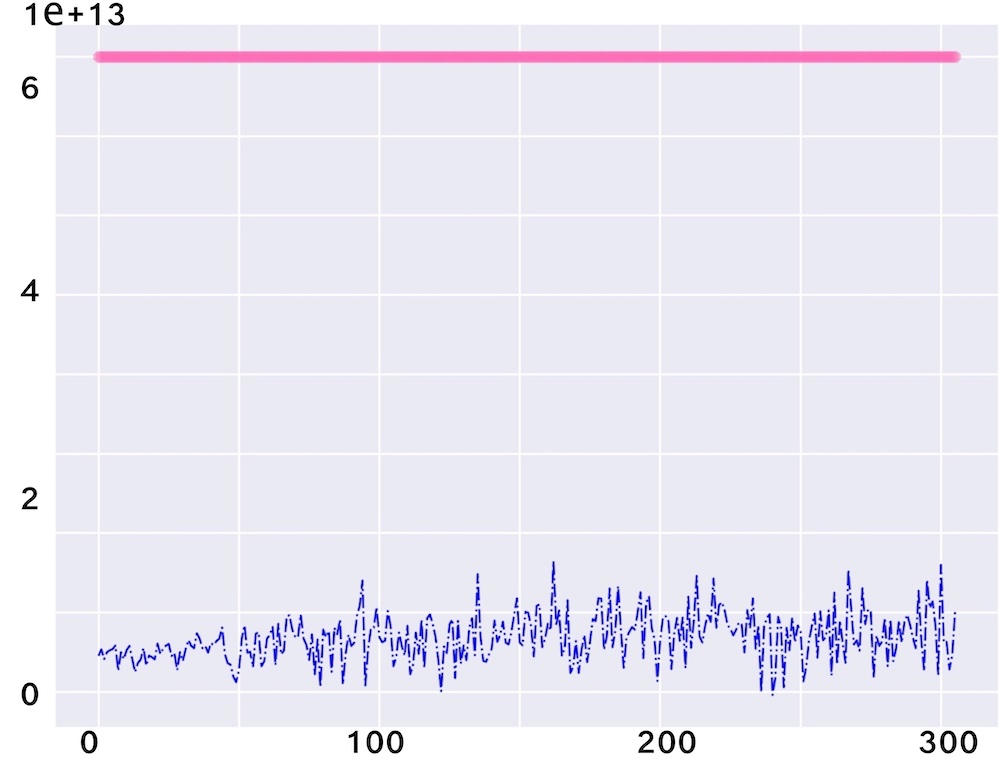}}
\subfigure[SVR(Poly)]{\includegraphics[width=1.36in]{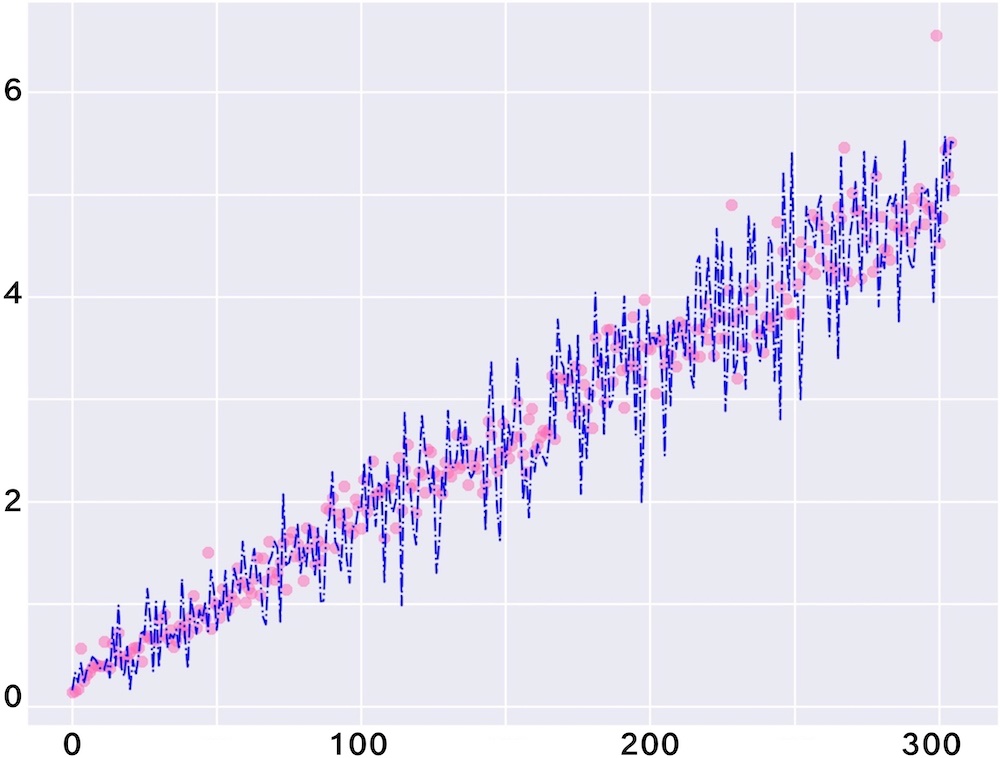}}
\subfigure[SVR(RBF)]{\includegraphics[width=1.36in]{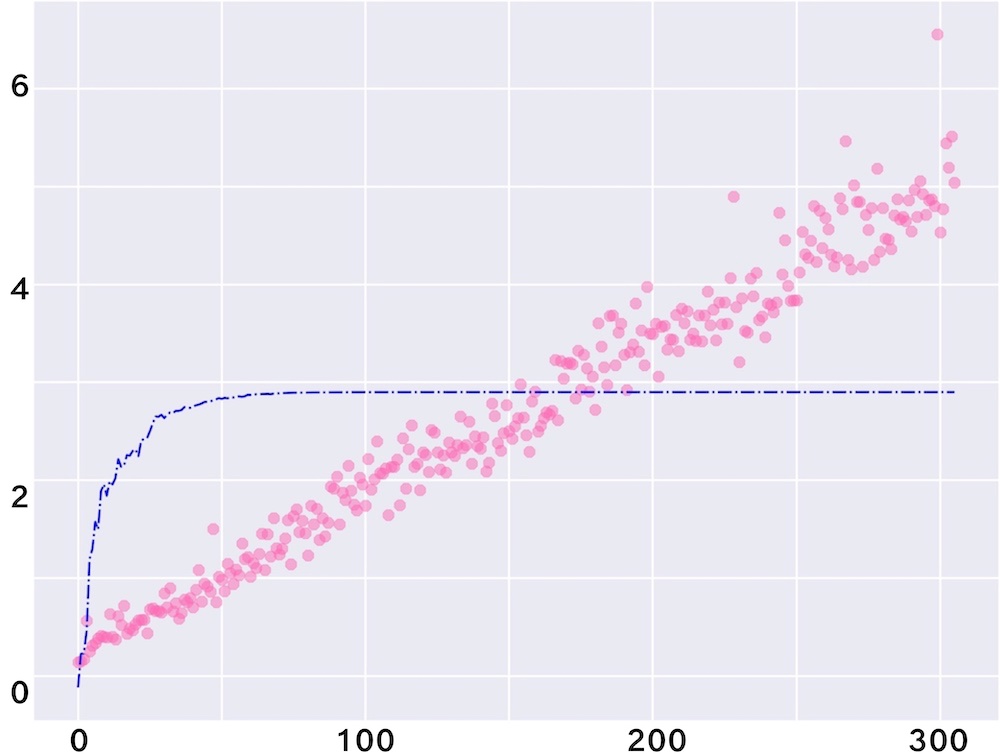}}
\subfigure[Theil]{\includegraphics[width=1.36in]{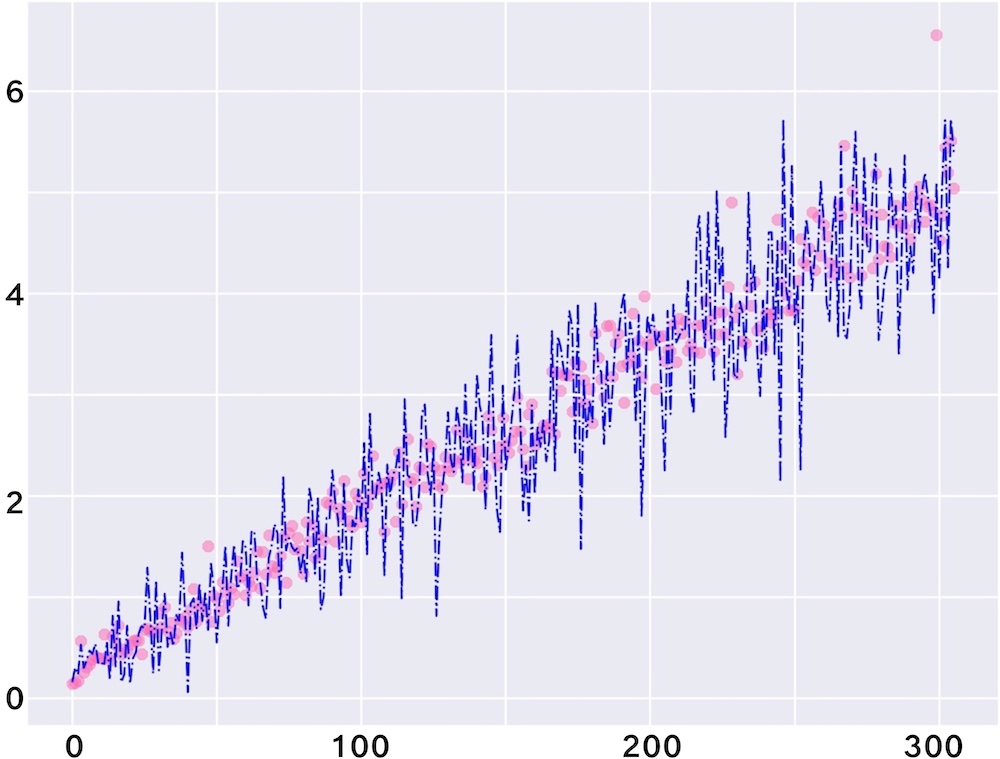}}
\subfigure[ICNet-NN]{\includegraphics[width=1.36in]{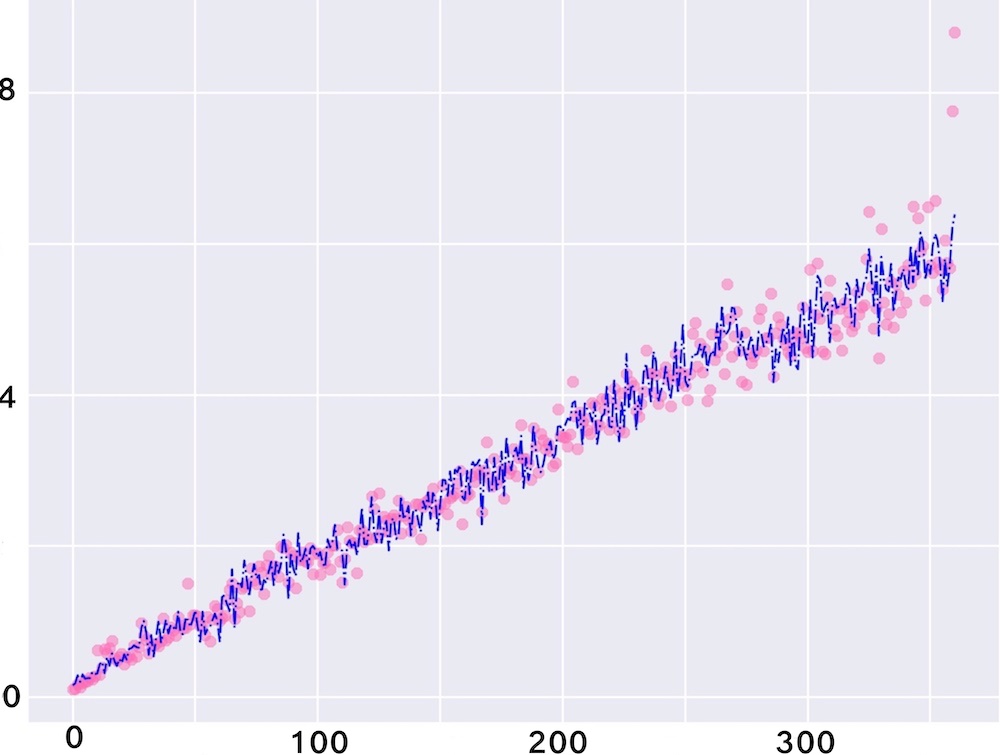}}
\caption{Comparison between predictions and real values: Pink dot are real values, blue lines are the predictions. x-axis is data index in testing data while y-axis is runtime value in log scale. Note that only SGD has different y-axis scale, i.e., 1e+13.}
\vspace{-20pt}
\label{prediction_behave}
\end{figure*}
Next, Fig. \ref{prediction_behave} illustrates several predicted value along with real value to analyze the prediction characterization.
Since there is little difference in dataset 2, we choose several competitive baselines in dataset 1 experiments under all feature setting. Several baselines performed very badly such as OMP and SGD which only output values around a constant level. SVR (RBF) is also bad and yield constant value when the real runtime is larger than a threshold. The results of EN and LASSO is positively related to the real values, but the correlation parameters are significantly different from the truth. Linear, RR, SVR (POLY) and Theil predicted the values that are relatively closer than that of the other baselines, but with high variance.
\vspace{-15pt}
\subsection{Case Study: Attentions on Attributes}
The subsection studies the attention mechanism quantitatively. Several circuits are evaluated, as shown in Table \ref{case_study}. Gate number consistently attracted greater attention than the gate type by 9.64\% on average. This motivates us to study the correlation between actual runtime and gate number. The Pearson (P) and Spearman (S) correlation are 0.8238 and 0.9722 on average in Table \ref{case_study}. Take circuit c7553 as an example, the runtime is 2.37\% of gate number. Different circuits show different linear parameters, which gives us a convenient message that can accurately predict the deobfuscation time and could serve circuit obfuscation task.
\begin{table}[hbt]
    \centering
        \caption{Case study: attributes and extracted rules.}
    \label{case_study}
    \begin{tabular}{l|cc|c|c}
        \toprule \hline
circuit & gate \# &  gate type & corr(P/S) & linear param \\ \hline
c7553 & 56.40\% & 43.59\% & 0.8754 / 0.9345  & 0.0237 \\
c499  & 54.39\% & 47.05\% & 0.8149 / 0.9965  & 0.1300  \\
c2670 & 52.94\% & 47.05\% & 0.7769 / 0.9753  & 0.0559 \\
c1335 & 56.27\% & 43.72\% & 0.8282 / 0.9846  & 0.0599 \\
\bottomrule
    \end{tabular}
\end{table}

\section{conclusion} \label{sec:conclusion}
In this work, we have introduced a neural network model for recovering SAT runtime on ICs, which expedites the evaluation on the hardness of obfuscated instances and therefore boosts the efficiency of developing obfuscation policy. To properly fuse graph structure and gate features, an enhanced graph convolutional operator is introduced. The proposed ICNet can avoid attribute propagation which is in the original GCN but not suitable for ICs. ICNet automatically extracts determinant features and aggregates gate representation regarding the runtime. Experiments on real-world datasets suggest that the proposed model is capable of modelling the runtime regarding the circuit graph accurately and stably, improving the baselines by a significant margin.


